\documentclass{article}
\usepackage{spconf,amsmath,graphicx}

\usepackage[retainorgcmds]{IEEEtrantools}
\usepackage{amssymb}
\usepackage{accents}
\usepackage{multirow}
\usepackage{algorithmic}
\usepackage{algorithm}
\usepackage{amsthm}

\DeclareMathOperator{\minimize}{minimize}
\DeclareMathOperator{\maximize}{maximize}
\newtheorem{prop}{Proposition}

\title{Network Topology Adaptation and interference coordination for Energy Saving in Heterogeneous Networks}
\name{Quan Kuang$^\star$, Xiangbin Yu$^\dagger$, Wolfgang
  Utschick$^\star$ \thanks{The work of X.~Yu was supported by National
    Natural Science Foundation of China (61571225), and QingLan Project of
    JiangSu.} }

\address{
\begin{tabular}{cc}
$\star$Dept of Electrical and Computer Engineering &  $\dagger$ College of Electronic and Information Engineering \\
Technische Universit\"at M\"unchen & Nanjing University of Aeronautics and Astronautics\\
Germany, \{quan.kuang, utschick\}@tum.de & Nanjing, China, yxbxwy@nuaa.edu.cn
\end{tabular}
}

\begin{document}
\ninept
\maketitle
\begin{abstract}
  Interference coupling in heterogeneous networks introduces the
  inherent non-convexity to the network resource optimization problem,
  hindering the development of effective solutions. A new framework
  based on multi-pattern formulation has been proposed in this paper
  to study the energy efficient strategy for joint cell activation,
  user association and multicell multiuser channel allocation. One key
  feature of this interference pattern formulation is that the
  patterns remain fixed and independent of the optimization
  process. This creates a favorable opportunity for a linear
  programming formulation while still taking interference coupling
  into account.  A tailored algorithm is developed to solve the
  formulated network energy saving problem in the dual domain by
  exploiting the problem structure, which gives a significant
  complexity saving compared to using standard solvers. Numerical
  results show a huge improvement in energy saving achieved by the
  proposed scheme.
\end{abstract}
\section{Introduction}
\label{sec:intro}

The densification and expansion of wireless networks pose new
challenges on interference management and reducing energy
consumption. In a dense heterogeneous network (HetNet), base stations
(BSs) are typically deployed to satisfy the peak traffic volume and
they are expected to have low activity outside rush hours such as
nighttime. There is a high potential for energy saving if BSs can be
switched off according to the traffic load.

Obviously, cell activation is coupled with user association: the users
in the muted cells must be re-associated with other BSs. In addition,
cell muting and user re-association impose further challenges on
interference management, since the user may not be connected to the BS
with the strongest signal strength. This interference issue can be
resolved by interference coordination, i.e., properly sharing the
channels among multiple cells and then distributing them to the
associated users in each cell. Hence, to obtain energy-efficient
resource management strategies, multicell multiuser channel assignment
should be integrated into the optimization of the cell activation and
user association.

However, the resource management that considers the above elements
jointly is very challenging mathematically because the inter-cell
interference coupling leads to the inherent non-convexity in the
optimization problems. To make the problems tractable, the previous
studies relied on worst-case interference assumption
\cite{Pollakis2012,Cavalcante2014a}, average interference assumption
\cite{Son2011a,Kim2013a}, or neglecting inter-cell interference
\cite{Su2013}. In these works, the interference was assumed
\emph{static} (or absent), i.e., independent of the resource
allocation decisions in each cell, when estimating the user achievable
rate. Clearly, this is a suboptimal design because the BS deactivation
will cause interference fluctuation in the network, hence affecting
the user rate.


This paper is developing a new framework for energy-efficient resource
management to consider the interference coupling caused by cell
deactivation. The idea is to pre-calculate the user rate under each
possible \emph{interference pattern} (i.e. an interference scenario in
the network, described as one combination of ON/OFF activities of the
BSs), and then perform resource allocation among these patterns. This
allocation yields the actual interference and the corresponding user
achievable rates that well match the interference at the same time.


\section{System Model}
\label{sec:system-model}

Consider a downlink HetNet, where a number of small cells are embedded
in the conventional macro cellular network. The set of all (macro and
small) cells is denoted by $\mathcal{B} = \{1,2,\cdots,B \}$. The
cells can be switched on or off every time period $T$ (say, many
minutes). In this relatively long decision period, we adopt test
points as an abstract concept to represent demands of users
\cite{Cavalcante2014a}. The test points can be chosen from typical
user locations, or we can simply partition the geographic region into
pixels and then each pixel becomes one test point.  The set of test
points is denoted by $\mathcal{K} = \{1,2,,\cdots K\}$ (In our model,
each test point can represent multiple co-located users). The traffic
demand of each test point $k \in \mathcal{K}$ is represented by a
minimum required average rate $d_k$ during one period of $T$, which is
assumed known via traffic estimation algorithms.  We are interested in
developing adaptive strategies for every period of $T$ to accommodate
the traffic requirement with minimum network energy consumption,
taking into account the inter-cell interference coupling.

The enabling mechanism is to characterize the interference by
specifying the interference patterns, each of which defines a
particular ON/OFF combination of BSs. We use the pattern activity
vector $\mathbf{a}_i = (a_{i1}, a_{i2}, \cdots, a_{iB})^T$ to
indicate the ON/OFF activity of the BSs under pattern $i$, where
\begin{equation}\label{pattern_actVect}
  a_{ib} =    \left\{  \begin{array}{rl}
      1 & \text{if BS $b$ is ON under pattern $i$}   \\
      0 & \text{otherwise}
    \end{array} \right.
\end{equation}
We denote the set of pre-defined patterns by $\mathcal{I} =
\{1,2,\cdots, I\}$ and further define the matrix $\mathbf{A} =
(\mathbf{a}_1, \mathbf{a}_2,\cdots, \mathbf{a}_I)$ to combine the
activity vectors for all candidate patterns. In order to fully
characterize the interference scenarios in a network of $B$ cells,
generally speaking, $2^B$ patterns are needed. However, since BSs
with large distance have weak mutual interference, omitting some patterns
will not affect the accurate estimation of user achievable rates. We
will discuss more on this next (see Proposition 1 in Section
\ref{sec:rate-constr-energy}).

Based on the above pattern definition, we establish a framework to
optimize the cell activation, test point (user) association and
multicell multiuser channel assignment jointly.

Firstly, the multi-cell channel allocation is translated into
partitioning the spectrum across all patterns. In a slow timescale
considered in this paper, all frequency resources can be assumed to
have equal channel conditions. Denote the spectrum allocation profile
by $\boldsymbol{\pi} = (\pi_1, \ldots, \pi_i, \ldots, \pi_I)^T \in
\Pi$, where $\pi_i$ represents the fraction of the total bandwidth
allocated to pattern $i$ and $\Pi = \{\boldsymbol{\pi} :
\sum_{i\in\mathcal{I}} \pi_i = 1, \pi_i \geq 0, \forall i\}$. Then the
total bandwidth fraction allocated to BS $b$ is $\mathbf{A}(b,:)\times
\boldsymbol{\pi}$, where $\mathbf{A}(b,:)$ denotes the $b$-th row of
the matrix $\mathbf{A}$.

Secondly, denote by $\alpha_{kbi} \geq 0$ the fraction of resources
that BS $b$ allocates to test point $k$ under pattern $i$. Naturally,
each BS is allowed to use up to $\pi_i$ resources under pattern $i$
for its associated test points, expressed as $\sum_{k\in \mathcal{K}}
\alpha_{kbi} \leq \pi_i, \forall b, \forall i$. Note that the
association is implicitly indicated by $\alpha_{kbi}$, i.e.,
$\alpha_{kbi} > 0$ means test point $k$ is associated with BS $b$
under pattern $i$, while zero value of $\alpha_{kbi}$ means that they
are not connected. In this formulation, test point $k$ is allowed to
be connected to multiple BSs. This can be equivalently viewed as
multiple users co-located at the same test point, and each BS serves
one user individually. In this paper, we assume a single-user detector
at each receiver.

Finally, we define the usage of BS $b$ as $\rho_b=\sum_k \sum_i
\alpha_{kbi}$. The definition of $\alpha_{kbi}$ leads to $0 \leq
\rho_b \leq 1, \forall b \in \mathcal{B}$.

\subsection{Rate model}

Assuming flat power spectral density (PSD) of BS transmit power and
the noise, the received SINR of the link connecting BS
$b$ to test point $k$ under pattern $i$ is
\begin{equation}\label{SINR}
  \textrm{SINR}_{kbi} = \frac{a_{ib} P_b G_{bk}\|h_{bk,n}\|^2 }{\sigma^2 + \sum_{l\neq b} a_{il} P_l G_{lk}\|h_{lk,n}\|^2 }
\end{equation}
where $a_{ib}$ is the cell activation indicator as given in
(\ref{pattern_actVect}), $P_b$ is the PSD of BS $b$, $\sigma^2$ is the
received noise PSD. We denote the channel gain between BS $b$ and test
point $k$ over the $n$-th frequency resource by $\sqrt{G_{bk}}
h_{bk,n}$ where $G_{bk}$ is the large-scale coefficient including
antenna gain, path loss and shadowing, and $h_{bk,n}$ accounts for the
small-scale fading. We assume $\{h_{bk,n},\forall b, \forall k,
\forall n\}$ are independent and identically distributed
(i.i.d.). Hence, the ergodic rate of test point $k$ served by the $b$-th BS
under pattern $i$ can be written as
\begin{equation}\label{RatePerCarrier}
  \bar{r}_{kbi}  = \alpha_{kbi} \underbrace{ W
    \mathbb{E}_{\mathbf{h}} \left[ \log_2 \left(1+
        \textrm{SINR}_{kbi}\right) \right]}_{ \triangleq r_{kbi}}  = \alpha_{kbi} r_{kbi}
\end{equation}
where $W$ is the system bandwidth, $\mathbf{h} \triangleq
(h_{1k,n},h_{2k,n},\ldots, h_{Bk,n})$.

Finally, the total rate of test point $k$ is obtained by summing up
the contributions from all associated BSs and patterns, as
\begin{equation}\label{RateCombined}
  R_k  = \sum_{i \in \mathcal{I}} \sum_{b \in \mathcal{B}} \bar{r}_{kbi}= \sum_{i \in \mathcal{I}} \sum_{b \in \mathcal{B}} \alpha_{kbi} r_{kbi}.
\end{equation}
Note that $r_{kbi}$ can be pre-calculted using (\ref{RatePerCarrier})
and hence treated as constants during the optimization.

\subsection{Energy consumption model}
As mentioned previously, the BS usage vector is defined as
$\boldsymbol{\rho} = (\rho_1, \cdots, \rho_B)^T$, where $\rho_b =
\sum_k\sum_i \alpha_{kbi}$. A typical power consumption model for BSs
consists of two types of power consumption: fixed power consumption
and dynamic power consumption that is proportional to BS's utilization
\cite{Son2011a}. Denote by $P_b^{\text{OP}}$ the maximum
operational power of BS $b$ if it is fully utilized (i.e., $\rho_b =
1$), which includes power consumption for transmit antennas as well as
power amplifier, cooling equipment and so on. We can then express the
total power consumption by all BSs as
\begin{IEEEeqnarray}{rCl}
  \label{eq:19}
P^{\text{tot}} =  \sum_{b\in \mathcal{B}} \left[ (1-q_b)\rho_b P_b^\textrm{OP} +
    q_b |\rho_b|_0 P_b^\textrm{OP} \right]
\end{IEEEeqnarray}
where $q_b \in [0,1]$ is the portion of the fixed power
consumption for BS $b$ as long as it is switched on, and $|x|_0$ is the
function that takes the value of 0 if $x=0$ or the value 1
otherwise. Note that by setting $q_b = 1$ we arrive at a constant
energy consumption model considered in
\cite{Pollakis2012,Niu2010}, which is a reasonable assumption for
macro BSs. However, the small BSs such as pico or femto BSs may have smaller
value of $q_b$ because they do not usually have a big power amplifier
or cooling equipment.

\section{Rate-constrained energy saving}
\label{sec:rate-constr-energy}

\subsection{Problem formulation}
\label{sec:problem-formulation}

The joint optimization of cell activation, user association and
interference coordination for energy
saving can be formulated as
\begin{IEEEeqnarray}{rCl}\label{P1_rateConstrained}
  \displaystyle\mathop{\minimize}_{\boldsymbol{\alpha},\boldsymbol{\pi}}
  \quad
  &&    P^{\text{tot}}= \sum_{b\in \mathcal{B}} \left[ (1-q_b)\rho_b P_b^\textrm{OP} +
    q_b |\rho_b|_0 P_b^\textrm{OP} \right]  \IEEEyesnumber\IEEEyessubnumber\label{obj}\\
  \text{subject to} \quad && \rho_b =  \sum_{k\in\mathcal{K}}
  \sum_{i\in\mathcal{I}} \alpha_{kbi}, \  \forall b \IEEEyessubnumber \label{cons_rho}\\
  && \sum_{i \in \mathcal{I}} \sum_{b \in \mathcal{B}}  \alpha_{kbi} r_{kbi} \geq d_k, \ \forall k  \IEEEyessubnumber\label{conQoS}\\
  && \sum_{k \in \mathcal{K}} \alpha_{kbi} \leq \pi_i, \forall b,\ \forall i   \IEEEyessubnumber \label{const_BS allo}\\
  && \sum_{i \in \mathcal{I}} \pi_i = 1  \IEEEyessubnumber\label{cons_pi}\\
  && \pi_i \geq 0, \ \forall i, \quad \alpha_{kbi} \geq 0, \ \forall
  k,b,i  \IEEEyessubnumber\label{con_nonnegative}
\end{IEEEeqnarray}
where \eqref{conQoS} specifies the traffic demand of all test points, and
all variables and parameters have been explained in Section
\ref{sec:system-model}.

The difficulty of solving \eqref{P1_rateConstrained} lies in two
facts. The first is the combinatorial objective function involving the
$\ell_0$-norm. The other is that the number of all possible patterns
in the network grows exponentially with the number of cells as $2^B$,
resulting in huge problem dimension for a moderate-sized
network. Fortunately, the following Proposition \ref{prop0} identifies
that only a small number of patterns out of $2^B$ are needed for
resource allocation to achieve the optimality.

\begin{prop}\label{prop0}
  There exists an optimal solution to problem
  \eqref{P1_rateConstrained} that activates at most $K+B+1$ patterns,
  i.e., $|\{i\in \mathcal{I} : \pi_i > 0 \}| \leq K+B+1$.
\end{prop}



The proof is provided in the appendix. The sparsity
structure identified by this proposition is exploited for the proposed
dual cutting plane method to reduce the computational complexity (see
Section \ref{sec:complexity-algorithm} for more discussion).

\subsection{Solving network energy saving problem}
\label{sec:solv-energy-saving}
We now turn the attention to solving (\ref{P1_rateConstrained}) assuming it
is feasible.
The idea is to apply
\emph{reweighted $\ell_1$-norm minimization methods}
\cite{Candes2008}, originally proposed to enhance the data acquisition
in compressed sensing. It is known that for nonnegative scalar $x \geq
0$, $|x|_0 = \lim_{\epsilon \rightarrow 0}
\frac{\log(1+x\epsilon^{-1})}{\log(1+\epsilon^{-1})}$
\cite{Pollakis2012}. With a small design parameter $\epsilon > 0$, we
neglect the limit and then approximate the $\ell_0$-norm as
\begin{equation}\label{MM_obj}
  |x|_0 \approx \frac{\log(1+x\epsilon^{-1})}{\log(1+\epsilon^{-1})}.
\end{equation}
Relying on (\ref{MM_obj}) and ignoring unnecessary constants, the
problem (\ref{P1_rateConstrained}) can be approximately solved by the
following problem:
\begin{IEEEeqnarray}{rCl}\label{P_approx}
   \mathop{\minimize}_{(\boldsymbol{\alpha}, \boldsymbol{\pi}) \in
    \mathcal{X}}  \quad &&  f(\boldsymbol{\rho}) \triangleq \sum_{b\in
    \mathcal{B}}\left[ (1-q_b) P_b^{\textrm{OP}} \rho_b + \frac{q_b
      P_b^{\textrm{OP}} \log(\epsilon +\rho_b)}{\log(1+\epsilon^{-1})}  \right] \IEEEnonumber  \\
  \textrm{subject to} \quad && \rho_b = \sum_{k\in\mathcal{K}}
  \sum_{i\in\mathcal{I}} \alpha_{kbi}, \  \forall b  \IEEEnonumber \\
  && \sum_{i \in \mathcal{I}} \sum_{b \in \mathcal{B}} \alpha_{kbi}
  r_{kbi} \geq d_k, \ \forall k. \IEEEyesnumber
 \end{IEEEeqnarray}
where $\mathcal{X}$ is defined by (\ref{const_BS allo}),
(\ref{cons_pi}) and (\ref{con_nonnegative}).

 Note that (\ref{P_approx}) is a continuous problem unlike the one in
 (\ref{P1_rateConstrained}) involving combinatorial terms. However, problem
 (\ref{P_approx}) is not convex since it minimizes a concave
 function. Fortunately, it falls into the framework of
 difference-of-convex (DC) functions and therefore can be efficiently
 solved by the convex-concave procedure \cite{Kuang2012}.

 Specifically, by applying the first-order Taylor expansion to the
 objective function in (\ref{P_approx}) at the point
 $\boldsymbol{\rho}^{(t-1)}$ obtained in $(t-1)$-th iteration, we
 arrive at the following problem for the $t$-th iteration:
\begin{IEEEeqnarray}{rCl}\label{P_rateConst_L1_reweighted}
\IEEEyesnumber\IEEEyessubnumber*
  \mathop{\minimize}_{(\boldsymbol{\alpha}, \boldsymbol{\pi}) \in
    \mathcal{X}}  \quad &&
  \sum_{b\in\mathcal{B}} w_b^{(t)} \sum_{k\in\mathcal{K}}\sum_{i\in \mathcal{I}}
  \alpha_{kbi}   \\
  \textrm{subject to}
  \quad 
  &&  \sum_{i \in \mathcal{I}} \sum_{b \in \mathcal{B}}  \alpha_{kbi} r_{kbi} \geq d_k, \ \forall k   \label{consDualrate}
  \end{IEEEeqnarray}
where
\begin{equation}\label{eq_weight}
  w_b^{(t)} = (1-q_b) P_b^{\textrm{OP}} + \frac{q_b P_b^{\textrm{OP}}}{\log(1+\epsilon^{-1}) (\epsilon + \rho_b^{(t-1)} )}
\end{equation}
with
 \begin{equation}\label{eq_cellLoad}
   \rho_b^{(t-1)}= \sum_{k\in\mathcal{K}}
  \sum_{i\in\mathcal{I}} \alpha_{kbi}^{(t-1)}.
 \end{equation}

 It can be shown, by applying the results in \cite{Lanckriet2009},
 that any limiting point of $(\boldsymbol{\alpha}^{(t)},
 \boldsymbol{\pi}^{(t)})$ generated by the above convex-concave
 procedure as $t \rightarrow \infty$ is a stationary point of the
 problem \eqref{P_approx}. In practice, the reweighted $\ell_1$ method
 converges typically within 6-10 iterations and the largest
 improvement in sparsity is obtained in the first few iterations.


 Problem \eqref{P_rateConst_L1_reweighted} is a linear program. It can
 be efficiently solved by, e.g., interior-point methods, if the
 problem dimension $\mathcal{O}(IKB)$ is small. However, it is also
 desirable to solve \eqref{P_rateConst_L1_reweighted} by involving a
 large number of patterns. This could happen when we consider all
 possible $2^B$ patterns in order to calculate an optimal performance
 benchmark in a reasonable-sized network, or when the pre-selection
 still results in lots of candidate patterns for a large-scale
 network.  In such case, the state-of-the-art interior-point solvers
 cannot be applied, since they typically have cubic computational
 complexity in the problem dimension \cite{Nesterov1994}.
 Fortunately, the problem has an interesting structure that
 facilitates a tailored cutting plane method to solve the dual
 problem.


 By dualizing the constraint of (\ref{consDualrate}), we
can express the dual function as
\begin{equation}\label{P_rateConst_Dualfun}
  h(\boldsymbol{\mu}) = \inf_{(\boldsymbol{\alpha}, \boldsymbol{\pi})
    \in \mathcal{X}} \left\{  \sum_{k,b,i}
     \alpha_{kbi} w_b^{(t)} -\sum_{k,b,i}  \alpha_{kbi}r_{kbi} \mu_k   +\sum_k d_k \mu_k   \right\}
\end{equation}
where $\boldsymbol{\mu} = (\mu_1,\cdots, \mu_K)^T$ is the Lagrangian
multiplier. The corresponding dual problem can be stated as
\begin{equation}
  \label{eq:dualProblem}
  \mathop{\maximize}_{\boldsymbol{\mu}>0} \quad  h(\boldsymbol{\mu}).
\end{equation}

Since the strong duality holds for this linear program, the original
primal problem \eqref{P_rateConst_L1_reweighted} can be alternatively
solved by the dual problem \eqref{eq:dualProblem}. Following the idea
of cutting plane methods, we formulate a master problem as
 \begin{IEEEeqnarray}{rCl}\label{dualprob_energy_master}
\IEEEyesnumber\IEEEyessubnumber*
  \mathop{\maximize}_{\boldsymbol{\mu}>0, z}  \quad &&  z  \\
  \textrm{subject to} \quad   && \sum_{k,b,i}\alpha_{kbi}^{(t,j)}
  w_b^{(t)} -\sum_{k,b,i}  \alpha_{kbi}^{(t,j)} r_{kbi} \mu_k
  +\sum_k d_k \mu_k  \geq z, \IEEEnonumber  \\
 && \qquad \qquad \qquad \qquad \qquad \ \forall j\in \{0,\cdots, l-1\}  \IEEEyessubnumber*\label{consmaster}
\end{IEEEeqnarray}
and an inner problem as
\begin{equation}\label{dualprob_energy_inner}
  \mathop{\minimize}_{(\boldsymbol{\alpha}, \boldsymbol{\pi}) \in
    \mathcal{X}} \quad \sum_{k,b,i} \alpha_{kbi}  \left(w_b^{(t)} -  r_{kbi}  \mu_k^{(l)} \right)  +\sum_k d_k \mu_k^{(l)}
\end{equation}
respectively, where we denote the solution to
(\ref{dualprob_energy_master}) by $(\boldsymbol{\mu}^{(l)}, z^{(l)})$
and the solution to (\ref{dualprob_energy_inner}) by
$(\boldsymbol{\alpha}^{(t,l)}, \boldsymbol{\pi}^{(t,l)})$. The master
problem (\ref{dualprob_energy_master}) is refined for the next
iteration by adding $(\boldsymbol{\alpha}^{(t,l)},
\boldsymbol{\pi}^{(t,l)})$ to the constraint (\ref{consmaster}). In
this way, we iteratively solve (\ref{dualprob_energy_master}) and
(\ref{dualprob_energy_inner}) until $h(\boldsymbol{\mu}^{(l)}) \geq
z^{(l)}$, implying that we have solved the problem
(\ref{P_rateConst_L1_reweighted}) in the dual domain.

The difficulty with huge dimension has now been encapsulated in
problem \eqref{dualprob_energy_inner} and nicely resolved thanks to
the following Proposition \ref{prop2}. The master problem
\eqref{dualprob_energy_master} is a linear problem with small
dimension (not involving $2^B$ term) that can be trivially solved
using any standard solver.

\begin{prop}\label{prop2}
  The problem \eqref{dualprob_energy_inner} has a closed-form solution
  that can be expressed as
  \begin{equation}\label{eq:3}
    \alpha_{kbi}^{(t,l)} =    \left\{  \begin{array}{rl}
        1 & \text{if} \ i=\bar{i}, k=\bar{k}(b, \bar{i}), \
        \text{and}\  \tilde{r}_{kbi} < 0  \\
        0 & \text{otherwise}
      \end{array} \right.
  \end{equation}
  and
  \begin{equation}\label{eq:4}
    \pi_i^{(t,l)} =    \left\{  \begin{array}{rl}
        1 & \text{if} \ i=\bar{i}  \\
        0 & \text{otherwise}
      \end{array} \right.
  \end{equation}
where $\bar{k}(b,i)=\arg \min_k \tilde{r}_{kbi}$ with $\tilde{r}_{kbi}
=  w_b^{(t)} - r_{kbi}\mu_k^{(l)}$, and $\bar{i} =
\arg \min_i  \sum_b\left[
  \tilde{r}_{\bar{k}(b,i)bi}\right]_{-} $, where $[x]_{-} =\min(0,x)$.
\end{prop}

After solving \eqref{eq:dualProblem} by the cutting plane, the primal
solution can be found as \cite[Ch.6]{Bazaraa2013}: $
\boldsymbol{\alpha}^{(t)} = \sum_{j=0}^{l-1}\kappa_j
\boldsymbol{\alpha}^{(t,j)}$ and $ \boldsymbol{\pi}^{(t)} =
\sum_{j=0}^{l-1}\kappa_j \boldsymbol{\pi}^{(t,j)}$, where
$\{\kappa_j\}$ are the dual variables corresponding to the inequality
constraints of \eqref{consmaster}, which are available if we solve the
problem \eqref{dualprob_energy_master} by off-the-shelf interior-point
solvers.

Finally, the outermost iteration is to adjust the weights according to
(\ref{eq_weight}) and (\ref{eq_cellLoad}) and then the problem
(\ref{P_rateConst_L1_reweighted}) is solved again with the new
weights until convergence.

\subsection{Complexity of the algorithm}
\label{sec:complexity-algorithm}

If problem \eqref{P_rateConst_L1_reweighted} considering all possible
pattern is directly solved by interior-point methods, the complexity
is roughly $\mathcal{O}(I^3K^3B^3)$. By contrast, every iteration of
the proposed dual algorithm requires finding a solution to
\eqref{dualprob_energy_inner} by Proposition \ref{prop2}, and a
solution to \eqref{dualprob_energy_master} by interior-point
solvers. Specifically, solving \eqref{dualprob_energy_inner} requires
$\mathcal{O}(IKB)$, while the complexity of solving
\eqref{dualprob_energy_master} depends on the number of constraints in
\eqref{consmaster}, which is increased by one inequality per
iteration. Our numerical experiment suggests that the number of iterations
is roughly proportional to $K$. (This can be explained by the inherent
sparsity structure of the solution identified by Proposition 1. Since
the proposed algorithm activates one pattern per iteration (see
\eqref{eq:4}), the number of iteration is unsurprisingly much lower
than $I$ if $I$ is large). Consequently, it is safe to bound the
complexity of solving \eqref{dualprob_energy_master} as
$\mathcal{O}(K^3)$ per iteration. Hence, the overall complexity of the
proposed algorithm for solving \eqref{P_rateConst_L1_reweighted} is
$\mathcal{O}(IK^2B + K^4)$, much smaller than directly applying
interior-point solvers to \eqref{P_rateConst_L1_reweighted}. In Table
1, we report the algorithm running time for a network consisting of
$50$ users and $15$ cells, where the proposed algorithm outperforms a
commercial solver (Gurobi \cite{Gurobi} with the barrier method), as
$I$ increases.

\begin{table}[t]
\footnotesize
\renewcommand{\arraystretch}{1.0}
\caption{Algorithm running time. }
\label{tab:runTime}
\centering
\begin{tabular}{|c | c | c | c |c | }
\hline
Number of candidate patterns &  $19$  &  $2^6$   &  $2^9$  &  $2^{15}$  \\
\hline
\hline
Proposed algorithm (sec) & 4.2   & 10.2 & 13.8 & 313  \\
\hline
Interior-point solver (sec)  &  0.3 & 1.2  & 19.6 & 6346 \\
\hline 
\end{tabular}
\end{table}

\subsection{Initialization}

The cutting plane method should be initialized with a strictly primal
feasible solution in terms of \eqref{consDualrate}, otherwise the master
problem will become unbounded in the first iteration. We can solve the
following rate balancing problem to test the feasibility of
\eqref{P1_rateConstrained} and obtain a strictly primal feasible
solution if the original problem is feasible:
\begin{IEEEeqnarray}{rCl}\label{P_rateBalancing}
  \IEEEyesnumber\IEEEyessubnumber*
  \displaystyle\mathop{\minimize}_{(\boldsymbol{\alpha},\boldsymbol{\pi})\in
    \mathcal{X}, R_{\textrm{sum}}} \quad
  &&    - R_{\textrm{sum}}   \\
  \text{subject to} \quad && \beta_k R_{\textrm{sum}} - \sum_{i \in
    \mathcal{I}} \sum_{b \in \mathcal{B}} \alpha_{kbi} r_{kbi} \leq 0
  , \ \forall k \label{con_rateBal_1}
\end{IEEEeqnarray}
where $\beta_k = d_k / \sum_{k\in \mathcal{K}} d_k$.  Note that
problem \eqref{P_rateBalancing} is always feasible. We can again apply
cutting plane method to solve it, but without worrying about the
initialization (since we can always decrease $R_{\text{sum}}$ to make
sure \eqref{con_rateBal_1} is strictly satisfied).

\section{Numerical results and discussions}
\label{sec:results}

\subsection{Simulation setup}
\label{sec:simulation-setup}

\begin{figure}[!t]
\centering
\includegraphics[width=0.33\textwidth]{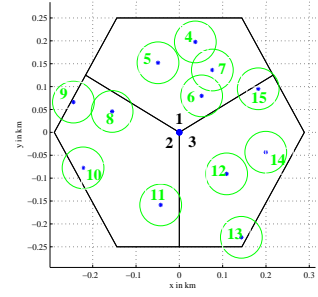}
\caption{A heterogeneous network consisting of 15 cells. }\label{fig_network}
\end{figure}

We consider a network consisting of 3 macro cells, each of which
contains 4 randomly dropped pico cells as shown in
Fig.~\ref{fig_network}.

The parameters for propagation modeling and simulations follow the
suggestions in 3GPP evaluation methodology, and summarized in Table I
of \cite{KuangAugust26-292014}. Based on the linear relationship between
transmit power and operational power consumption\footnote{We adopted
  the linear model in \cite{Fehske2009}: $P_b^\text{OP} =\alpha_b P_b
  +\beta_b $, where $P_b$ is the transmit power for BS $b$, $\alpha_b=
  \frac{22.6}{3}$ and $\beta_b = \frac{412.4}{3} $W if $b$ is a macro;
  otherwise $\alpha_b = 5.5$ and $\beta_b = 32 $W if $b$ is a pico.},
we calculate the maximum operational power $P^\text{OP}$ as 439W and
38W for macro and pico BSs, respectively. We further assume each macro
BS has a constant power consumption, i.e., $q_b =1$, $\forall b \in
\mathcal{B}_\text{macro}$, and the fixed power consumption of a pico
takes $50\%$ of the maximum operational power, i.e. $q_b = 0.5$,
$\forall b \in \mathcal{B}_\text{pico}$. Note that these assumptions
are made for providing concrete numerical results, and they are not
from the restriction of our formulation.

\subsection{Performance comparison}
\label{sec:perf-prop-scheme}

The baseline strategy in comparison is the energy saving optimization
scheme proposed by \cite{Cavalcante2014a}, where worst-case estimates
of the user rates resulted from no intercell interference coordination
are used to calculate the QoS requirements. This scheme can be cast
into the proposed framework by restricting the candidate pattern to a
single \emph{Reuse}-1 pattern.

\begin{figure}[!t]
\centering
\includegraphics[width=0.33\textwidth]{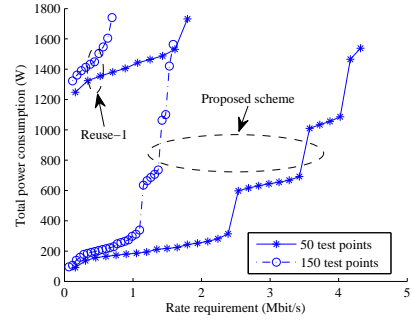}
\caption{A comparison of network power consumption between the proposed
  scheme and Reuse-1 scheme in \cite{Cavalcante2014a}. }\label{fig_compare}
\end{figure}

Fig.\ref{fig_compare} plots the network power consumption versus the
rate requirement of the test points, where $50$ and $150$ test points
are uniformly distributed within the network, and all test points are
assumed to have the same requirement for simplicity.

As shown, the network power consumption increases with the user rate
requirement for both schemes, but the proposed scheme has a
significantly power saving compared to the existing Reuse-1
scheme. For example, to satisfy $1$Mbit/s for 50 test points, the
proposed scheme only consumes 200W, whereas the Reuse-1 needs more
than 1400W. Moreover, the maximum rate requirement that the network
can support has been greatly improved by the proposed scheme. We
observe, for example, the maximum feasible rate in 50-test-point case
increases from $1.8$Mbit/s to $4.3$Mbit/s by using the proposed
scheme. The performance gains of the proposed strategy come from its
ability to manage the interference by resource allocation and
explicitly take into account the interference coupling caused by cell
(de)activation when estimating the user rate.


\section{Appendix}

In this appendix, the proof of Proposition \ref{prop0} is provided.
By letting $\alpha_{kbi} = \pi_i\theta_{kbi}$, the original problem
can be equivalently rewritten as
\begin{IEEEeqnarray}{rCl}\label{P1_rateConstrained_changeVairable}
  \displaystyle\mathop{\minimize}_{\boldsymbol{\theta},\boldsymbol{\pi}}
  \quad
  &&    P^{\text{tot}}= \sum_{b\in \mathcal{B}} \left[ (1-q_b)\rho_b P_b^\textrm{OP} +
    q_b |\rho_b|_0 P_b^\textrm{OP} \right]  \IEEEyesnumber\IEEEyessubnumber*\label{prop1_obj} \\
  \text{subject to} \quad && \rho_b =  \sum_{i\in\mathcal{I}} \pi_i
  \sum_{k\in\mathcal{K}} \theta_{kbi}, \  \forall b \label{prop1_cons_rho} \\
  && \sum_{i \in \mathcal{I}} \pi_i \sum_{b \in \mathcal{B}}  \theta_{kbi} r_{kbi} \geq d_k, \ \forall k  \label{prop1_conQoS} \\
  && \sum_{k \in \mathcal{K}} \theta_{kbi} \leq 1, \forall b,\ \forall i    \label{prop1_const_BS allo}\\
  && \sum_{i \in \mathcal{I}} \pi_i = 1  \label{prop1_cons_pi} \\
  && \pi_i \geq 0, \ \forall i, \quad \theta_{kbi} \geq 0, \ \forall
  k,b,i  \label{prop1_con_nonnegative}
\end{IEEEeqnarray}
In the following, we show that if an optimal solution
$(\boldsymbol{\theta}^\star, \boldsymbol{\pi}^\star)$ exists we can
then obtain the same optimal objective with
$(\boldsymbol{\theta}^\star, \boldsymbol{\pi'})$ where
$\boldsymbol{\pi'}$ only has $K+B+1$ nonzero entries out of $
|\mathcal{I}|$ entries.

We first define $\mathbf{t}_i=(t_{1i},\cdots,t_{bi},\cdots, t_{Bi})^T$
with $t_{bi} = \sum_{k\in\mathcal{K}} \theta_{kbi}^\star$, and
$\mathbf{R}_i = (R_{1i},\cdots, R_{ki}, \cdots$, $R_{Ki})^T$ with
$R_{ki} = \sum_{b \in \mathcal{B}} \theta_{kbi}^\star r_{kbi}$. Then
define $\boldsymbol{\rho} = (\rho_1,\cdots, \rho_B)^T$ and $\mathbf{d}
= (d_1, \cdots, d_K)^T$. According to \eqref{prop1_cons_rho} and
\eqref{prop1_conQoS} (note that \eqref{prop1_conQoS} must achieve
equality at the optimum, otherwise the objective in \eqref{prop1_obj}
can be further reduced), the vector $(\boldsymbol{\rho}^T,
\mathbf{d}^T)^T = \sum_i \pi_i (\mathbf{t}_i^T, \mathbf{R}_i^T)^T$,
i.e., a convex combination of vectors $(\mathbf{t}_i^T,
\mathbf{R}_i^T)^T, \forall i\in \mathcal{I}$, with $\pi_i$ as
coefficients. By Caratheodory's Theorem, $(\boldsymbol{\rho}^T,
\mathbf{d}^T)^T$ can be represented by at most $K+B+1$ of those
vectors. Denoting the resulting coefficients by
$\boldsymbol{\pi}^\prime$, we prove the Proposition.

\vspace{10cm}


\end{document}